# Chemical energy in an introductory physics course for the life sciences

Benjamin W. Dreyfus, Julia Gouvea, Benjamin D. Geller,
Vashti Sawtelle, Chandra Turpen, and Edward F. Redish

*Department of Physics, University of Maryland, College Park, MD 20742*

**Abstract.** Energy is a complex idea that cuts across scientific disciplines. For life science students, an approach to energy that incorporates chemical bonds and chemical reactions is better equipped to meet the needs of life sciences students than a traditional introductory physics approach that focuses primarily on mechanical energy. We present a curricular sequence, or thread, designed to build up students' understanding of chemical energy in an introductory physics course for the life sciences. This thread is designed to connect ideas about energy from physics, biology, and chemistry. We describe the kinds of connections among energetic concepts that we intended to develop to build interdisciplinary coherence, and present some examples of curriculum materials and student data that illustrate our approach.

## I. INTRODUCTION

Energy is a central concept in all of the scientific disciplines, universally useful for describing and explaining a range of phenomena.[1] However, energetic frameworks are applied variably across the science disciplines, each utilizing aspects of the concept most relevant to the phenomena of interest: falling objects, chemical reactions, or ecosystem dynamics. In science instruction, different disciplines tend to present these frameworks in isolation[2], which can make the teaching and learning of energy concepts appear fragmented rather than unified.[3]

To build connections between physics, biology, and chemistry[4], an interdisciplinary understanding of energy is necessary. The discipline-based education research literatures on energy largely fail to talk to each other across disciplinary boundaries[5], but these conversations become more essential as the sciences themselves become more interdisciplinary. Cooper and Klymkowsky[2] write "We are failing our students by not making explicit connections among the way energy is treated in physics, chemistry, and biology. We cannot hope to make energy a cross-cutting idea or a unifying theme until substantive changes are made to all our curricula." This paper presents one such approach to substantive curricular change that begins to make these explicit connections across disciplines.

In traditional introductory physics courses, the focus is on mechanical energy to the exclusion of other energy. If it is mentioned at all, "chemical energy" is treated as a black box, a "miscellaneous" form of energy whose role is to account for discrepancies when mechanical energy is not conserved, but it is not explored at a deeper level. Introductory physics for the life sciences (IPLS) courses are aimed at providing the tools to explain the physics principles that underlie complex phenomena in biology and chemistry. For students in the life sciences, there is a need to understand how chemical energy transformations at the molecular level connect with organism and ecosystem level flows. Because of the central role of chemical energy in biology, building a coherent framework of energy that connects physics to biology requires integrating ideas about chemical energy with the more canonical treatments of energy from physics. We conceptualize the concept of chemical energy as existing throughout the course as a recurring conceptual "thread." We describe our intentions in developing the chemical energy thread and present some examples of curriculum materials and student data that illustrate our approach.

In Section II, we provide background on the interdisciplinary course context in which our course materials were developed. In Section III, we explain the role of the chemical energy "thread" in the course and how it interacts with other threads. Section IV discusses the conceptual connections within the thread, and the motivations behind them. Section V describes some examples of the tasks that comprise the thread. In Section VI, we present some qualitative data illustrating preliminary student outcomes. While the course materials included in this paper are a limited selection, the full set of materials is freely available at the thread website.[6]

## II. BACKGROUND TO THE COURSE

The IPLS course[7] in which our materials are developed and used is part of the National Experiment in Undergraduate Science Education (NEXUS)[8], and represents the results of an interdisciplinary collaboration[9] bringing together perspectives from physics, biology, biophysics, chemistry, and education research. The NEXUS/Physics course is a two-semester course intended for life sciences majors. The course is structured as 150 minutes per week of lecture (using Peer Instruction and other interactive techniques) along with 1 hour of recitation (used for group problem solving) and 2 hours of lab.[10] While calculus is a formal prerequisite (as it is required anyway for biology majors at Maryland), the use of calculus in the chemical energy thread is primarily conceptual, and these materials could be used in an algebra-based course with little modification. The prerequisites also include a year of biology and a semester of chemistry, and therefore we design the curriculum in a way that builds on students' prior experiences in biology and chemistry coursework.[7]

This thread relies on highly simplified models of atomic and molecular interactions, in order to enable qualitative sense-making around much more complex processes and reactions that are discussed in the biology and chemistry courses. Therefore, there is no requirement for the physics instructor teaching this course to have sophisticated knowledge of chemistry or biology.

Our data from the course show that our students come in with ideas about energy from biology and chemistry. As an illustrative example, at the beginning of the energy unit in fall 2012, the professor asked the class "You talk about energy in your biology classes and your chemistry classes. So I want to know what you think energy is." Two students, Irene and Violet (all names are pseudonyms) simultaneously responded "ATP!" and then one cheered "Yeah!" They were talking about adenosine triphosphate, the molecule that Irene referred to in an interview as "the biological form of energy." The professor probed further about what they meant when they said that energy is ATP, and Sonia responded, "In biology it's the chemical bonds which hold energy." Sonia was using language about chemical bonds in a manner that is common in introductory biology courses[11], and which has been noted as problematic in the biology and chemistry education literature.[12–15] This brief episode makes it clear that teaching about energy to this student population, in a way that builds from their existing knowledge, must engage with chemical bonds and ATP, which are salient in these students' incoming understandings of energy.

### III. STRUCTURE OF THE CHEMICAL ENERGY THREAD

The NEXUS/Physics[7] course has multiple components: a wikibook with readings, interactive lectures with clicker questions, weekly group problem-solving sessions, homework problems, and labs. Chemical energy is included in the course as an instructional "thread" that runs through and links many aspects of the course and is not merely an independent unit. The goal of the chemical energy thread is to help students make stronger connections, both within physics and between physics and other disciplines. Conceptualizing the curriculum as threadlike has helped us support this goal in several ways.

Threads represent a structuring of the curriculum that builds expertise over time. Students need to encounter ideas and reasoning strategies many times in different contexts in order to develop expertise.[16] They don't have to "get it" in an all-or-nothing way the first time they see something. For example, we do not expect students to fully understand potential energy when they first grapple with it in the context of near-earth gravitational free-fall scenarios. Nor do we expect students to fully understand it when they engage in reasoning about charged particles interacting. We want to give students multiple opportunities to reason about potential energy across a variety of situations and support them in coordinating these understandings. A thread is more than just a conceptual sequence. It also must include opportunities for students to examine the links between concepts.

For this reason some of the problems and activities that comprise the thread are designed to ask students to explicitly consider the ways in which different ideas about energy are connected. Students are given multiple opportunities to connect chemical energy to other relevant descriptions of energy within physics (e.g. kinetic and potential energy; the relationship between energy and force), and to make connections among multiple ways of describing and representing energy (e.g. a focus on transfer of energy in and out of a system vs. a focus on energy transformations within a system vs. a focus on the energy of an object as a function of position), facilitating links to ideas about energy from chemistry and biology. Accomplishing all of this would be more difficult if chemical energy were simply added to the existing course as an isolated module.

Our curricular thread on chemical energy comprises a series of instructional tasks including clicker questions, homework problems, recitation group problem-solving activities, quiz and exam questions designed to help students develop coherence along the particular dimension of topical understanding of chemical energy. However, the tasks that constitute this thread are also components of other threads designed, for example, to develop productive epistemological stances or meta-representational competence. Figure 1 shows a few examples of how these threads intersect.

Our intention is for the multiple interacting threads to simultaneously work to develop different dimensions of scientific expertise. Attempting to influence one dimension of expertise may be facilitated by attention to other dimensions of scientific expertise. For example, developing a robust conceptual framework for ideas about energy can be facilitated by simultaneously developing the ability to understand and translate between different representational forms (graphical, diagrammatic, symbolic, and verbal). A third interacting component of expertise involves asking students to consider and evaluate differences in the way physics, chemistry, and biology leverage models of energy in order to make sense of different kinds of phenomena. This epistemological thread engages students in evaluating what they know and determining the realm of applicability for particular models of energy.

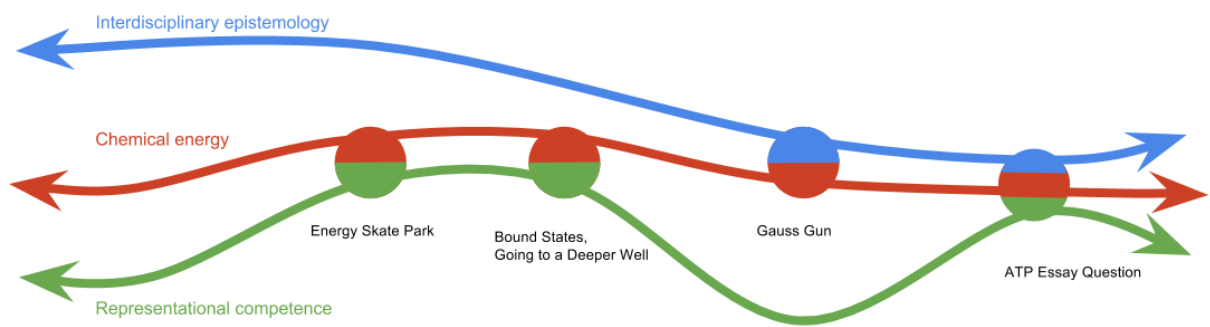

**FIG. 1.** A small section of the chemical energy thread and how it intersects with two other threads we are developing in the NEXUS/Physics course. The circles represent a few example tasks (homework, exam problems, etc.) that were designed to help students build up the ideas and connections in this thread. Split circles represent tasks that develop competence across multiple threads.

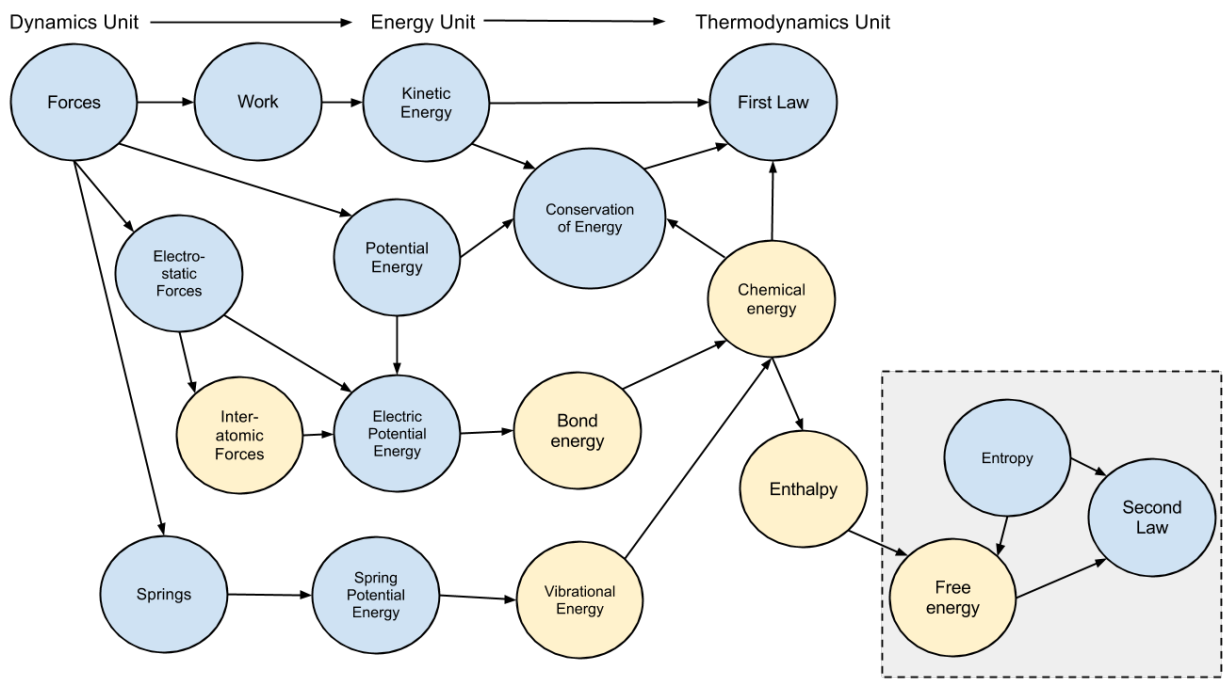

**FIG. 2.** The nodes represent conceptual components of the chemical energy thread, and the arrows represent links between these concepts. Blue nodes represent content typically included in introductory physics. Yellow nodes represent content added in service of building up an integrated treatment of chemical energy (see also Table 1).

# IV. CONTENT OF THE CHEMICAL ENERGY THREAD

Building across these curricular tasks to develop an understanding of chemical energy requires combining concepts traditionally covered in introductory physics courses with ideas that are more commonly taught in chemistry and biology. The nodes in Figure 2 represent the way we have built up the conceptual components of this thread in our course. Our thread asks students to explicitly reflect on the links between the canonical physics contexts and other disciplinary contexts. These important links are represented as the arrows in Figure 2. The chemical energy thread comprises a particular sequence of tasks that aims to support students in understanding and coordinating among these concepts. Our purpose in this paper is not to prescribe this particular sequence of tasks, but to articulate connections that should be scaffolded in developing a more complete model of chemical energy that will serve students across disciplinary contexts (Table 1). In this section we identify the connections that the chemical energy thread is intended to highlight, and in section V we discuss how the specific example tasks support building these connections.

| Thread Component | Motivation for inclusion |
| --- | --- |
| Introduce electrostatic forces in force unit | Emphasizes the forces that are most relevant at cellular and molecular scales, and sets the stage for electric potential energy |
| Include electric potential energy as one type of potential energy | Emphasizes that this energy is not fundamentally different from mechanical energy |
| Build up a model for chemical bonds using Lennard-Jones (L-J) potential | Models "chemical energy" associated with the formation and breaking of bonds in terms of potential and kinetic energies |
| Apply L-J model to chemical reactions | Links changes in chemical energy to changes in potential and kinetic energy at the molecular scale |
| Include chemical energy as component of internal energy | Connects First Law to chemical reactions |

**TABLE 1.** Selected content from the chemical energy thread, with the motivations for including it.

The thread starts at the very beginning of the course with the kinematics unit, which includes examples of motion at the microscopic scale. Students analyze the motion of cell-sized objects in homework and in lab[10], which establishes the idea that the models of mechanics in the course are valid at scales from macroscopic to molecular.[17] The specifics of these tasks are less relevant here than the general stage-setting for applying common reasoning across physical scales. The course moves some of the electrostatics material (traditionally covered in the second semester) to the first semester, to emphasize forces that are most relevant at cellular and molecular scales. The force unit introduces Coulomb's Law and electrostatic forces, including a careful treatment of charge polarization, showing how a neutral object can experience a net electric force as a result of the separation of charges, a crucial element in understanding atomic and molecular interactions.

When potential energy is introduced, electric potential energy is included as an integral part of the energy unit of the course rather than in a separate electricity unit, emphasizing that this energy is not fundamentally different from mechanical energy. This sets the stage for a model of chemical bond energy. To build up a mostly classical model for chemical bonds, we follow existing curricula[18–20] in using the Lennard-Jones potential[21] (Fig. 3a), which approximates the potential energy associated with the interaction of two atoms with an attractive term proportional to $1/r^6$ and a repulsive term proportional to $1/r^{12}$, where $r$ is the distance between the nuclei.

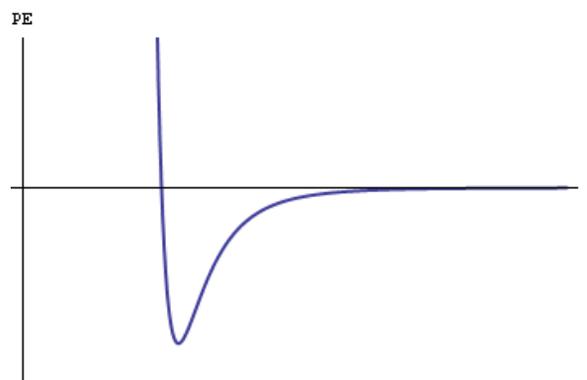

**FIG. 3a.** The Lennard-Jones potential, approximating the potential energy associated with the interaction of two atoms, as a function of the distance ($r$) between the atoms.

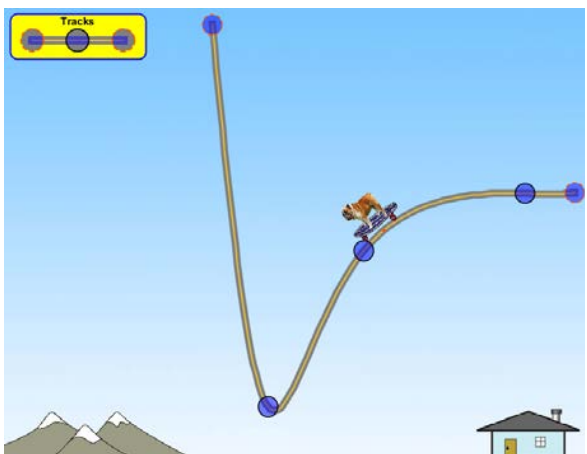

**FIG. 3b.** The *Energy Skate Park* simulation[22] is leveraged to draw analogies between chemical bonds and students' experience of gravitational potential energy.

The shape of the Lennard-Jones potential is justified to the students using primarily qualitative arguments. Building on traditional demonstrations like sticking a charged balloon to the wall[23], a charged particle can induce a dipole in a neutral atom. This leads to a force of attraction between the charged particle and the atom, though this attraction falls off more quickly with distance than the Coulomb force between two charges. Furthermore, a dipole (even a temporary dipole created by random fluctuations of the electron distribution in a neutral atom) can induce a dipole in another neutral atom and attract it, but this attraction is even weaker than the charge-dipole interaction. This is the Van der Waals force that our students have encountered in their chemistry classes. Without getting into the math, the $1/r^6$ dependence is plausible, since the Van der Waals attraction is many degrees weaker than the $1/r$ Coulomb potential except at very short distances. At large $r$, this attractive potential gives the expected qualitative result: the potential is relatively flat, indicating no significant interaction between neutral atoms at large separation. However, as $r$ decreases, this term suggests that the attraction continues to get stronger. Our students are familiar from chemistry with the Pauli exclusion principle, which prevents atoms from getting too close. Qualitatively, we expect this effective repulsion to be very strong at short distances (sufficient to overcome the $1/r^6$ attractive term) and to fall off quickly at longer distances (so that the attraction dominates), so a $1/r^{12}$ dependence is plausible.

Putting the two terms together, there is a minimum in the potential energy function that corresponds to an equilibrium (about which the system can oscillate if kinetic energy is present). The r at which this minimum occurs is the bond length for two bound atoms, and this bound state corresponds to a chemical bond. The relevant qualitative features of a chemical bond on which we want to focus emerge from this model: the bound state has a stable equilibrium[24] with negative potential energy (relative to the zero of potential energy set when the atoms are separated by a large distance), so an input of energy is needed to separate two bound atoms (i.e., to break the bond). Conversely, when two unbound atoms become bound, their potential energy decreases, and so conservation of energy dictates an increase of energy elsewhere ("energy is released"). Thus the "chemical energy" associated with the formation and breaking of bonds is explicitly modeled in terms of potential and kinetic energies. (For an example, see the "Bound states" task in Section V.)

The next step is to build up multiple bond-breaking and bond-formation events into a chemical reaction. A reaction is either exothermic or endothermic, depending on the overall sign after adding together all the energy changes associated with bond breaking and bond formation. These overall changes in "chemical energy" in chemical reactions are now linked to kinetic and potential energy at the molecular scale.

The thread extends later into the course as well, beyond the "energy" section. When the course moves into the laws of thermodynamics, it continues to include chemical energy among the types of energy that are considered. In a traditional introductory physics course, the First Law of Thermodynamics is used primarily in the context of ideal gases, and therefore the "internal energy" term is equated with thermal energy, energy that depends only on the temperature of the gas. In the NEXUS/Physics course, internal energy includes not only thermal energy but also chemical energy. Thus, changes in the internal energy of a system may be manifested not only as temperature changes but also as chemical reactions. This is more consistent with the First Law as it is taught in biology courses, where chemical reactions are central and temperature changes are not. This means that the total internal energy is undefined, since the total chemical energy (which includes potential energy) is undefined, and the zero of potential energy can be placed anywhere. This is a departure from the approach in a traditional introductory course, which may include an explicit expression for internal energy. However, this is not a problem, because only changes in internal energy have physical significance.

The chemical energy thread continues with links to enthalpy and free energy. Those topics are beyond the scope of this paper, but are discussed elsewhere.[25] Those links are essential to enabling students to make full connections to ideas about energy from their biology and chemistry courses, since biology and chemistry courses typically formulate reaction energies in terms of enthalpy (along with using Gibbs free energy to determine the spontaneity of reactions).

## V. EXAMPLE TASKS FOR STUDENTS

In this section, we present illustrative examples of the kinds of tasks and problems that comprise the thread and support our intention to make connections among concepts and among disciplines. These tasks are available on the NEXUS/Physics course website.[6]

Prior to any explicit instruction on chemical energy, but after the Work-Energy Theorem has been introduced, the students do a group problem-solving task on protein folding. This task has been discussed at length, including the process of revising it to support interdisciplinary learning, in ref. 26. The protein folding task asks students to reason about the relationship between force, work, and energy in the context of using optical tweezers to unfold proteins, and to connect these relationships from physics with the biological question of what it means for a protein to be in a stable state. Students come in with the idea from biology that a molecule at a lower energy state is "more stable," and so they are asked to coordinate this framework of energy as stability with the force/work/energy framework from mechanics. This task is intended to prime students for the rest of the chemical energy thread by having them think about force and energy on the scale of biomolecules, using the same physical principles that apply at the macroscopic scale. Question prompts include comparing a mutated and a wild-type protein, both in terms of stability and in terms of the work it would take to unfold them. Students are asked to use two different representations: a graph of energy vs. reaction coordinate[27] that represents the "energy landscape" of a folding protein, and a graph of force vs. extension that shows data from when a protein is stretched with an optical tweezer.[28]

Another group problem-solving task involves the *Energy Skate Park* simulation[22] from the PhET project. This simulation has a skateboarder on an editable track, and uses multiple representations to keep track of kinetic, (gravitational) potential, thermal, and total energy. The shape of the track itself doubles as a potential energy vs. position graph, since gravitational potential energy is proportional to height. The NEXUS/Physics course then uses *Energy Skate Park* as the foundation for a series of homework problems on chemical energy. An excerpt from one of these problems is given in Figure 4. In these problems, students use their physical intuitions about the relationships between energy, force, and motion, based on experience with gravity, and they extend this reasoning to cases where the relevant potential energy is not gravitational, but where a vertical location metaphor (e.g. "potential well"[29]) is still useful. Thus the skateboarder becomes an analogy for two interacting atoms, and the track is an analogy for their potential energy function (as in Fig. 3a and 3b).

B. Now suppose that the skateboarder starts *inside the well* at a zero velocity -- say at point x = -2.5 units with a total energy as shown by the heavy solid line.

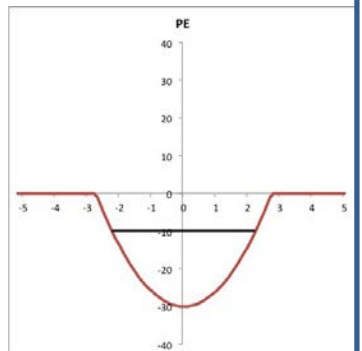

Describe the motion of the skateboarder and how her potential and kinetic energies change as she moves through the well.

C. Her total energy is shown is the figure as -10 units. How can this be? Is it reasonable for the total mechanical energy to be negative?

D. If she wants to climb out of the well and be at 0 kinetic energy at the point x = 3 units, how much energy would she need to gain?

E. The skateboarder is actually just an analogy for the cases we are interested in, which are interacting atoms. The potential energy of the interaction looks like the figure at the right.

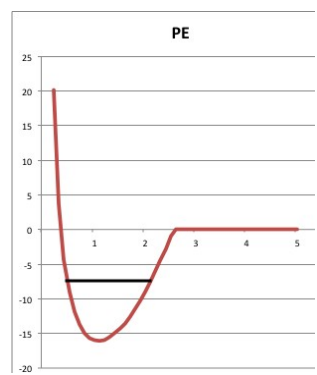

If the atoms have the energy of -7.5 units as shown by the solid line in the figure, describe their motion and how their potential and kinetic energies change as they move in the well.

F. If the atoms have an energy of -7.5 units as shown by the solid line in the figure, would you have to put energy in to separate the atoms or by separating them would you gain energy? How much? Explain why you think so.

**FIG. 4.** Excerpts from the "Bound states" problem.

Later in the series, the students are given a potential with multiple wells of different depths. This is used as an analogy for chemical reactions that involve going from one bound state to another, and helps students reconcile how it is that breaking a bond (such as in ATP[30]) can lead to the release of energy (because other stronger bonds are formed). Unlike the single well (as in Fig. 3a), where the horizontal axis represents the distance between two atoms, the multiple-well situation is more complicated, in that the independent variable on the graph does not correspond to a single physical parameter. While we recognize the limitations of this toy model (and encourage the students to explore these limitations), we believe that using this representation (among others) can be pedagogically useful because it provides a mechanical analogy that can help students bridge to their macroscopic intuitions, and because it can bridge to the reaction coordinate diagrams that students are familiar with from chemistry and biology courses (in which the reaction coordinate is also not rigorously defined in terms of physical parameters).

A demonstration and homework problem on the Gauss gun[31] make a similar point, asking students to reason about the Gauss gun as an analogy for an exothermic chemical reaction. The Gauss gun is a device consisting of a magnet and several metal spheres (Fig. 5). The sphere closest to the magnet (sphere 1 in the figure) is most strongly bound. When a new sphere (sphere 0 in the figure) is released from rest and sticks to the magnet, sphere 3 is ejected at high speed, so that the final kinetic energy of the system is greater than the initial kinetic energy. Students are asked, "Where did the energy come from?" This is a mechanical analog of an exothermic chemical reaction, in which a stronger bond is formed and a weaker bond is broken, resulting in the release of energy.

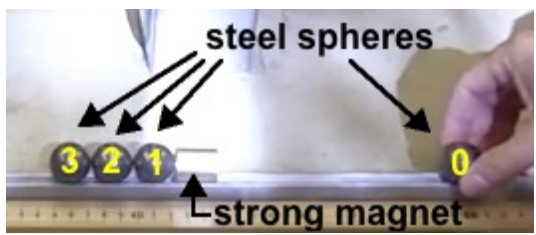

**FIG. 5.** The Gauss gun.

Several tasks then ask students to apply physical models for chemical energy to biological scenarios. In a homework problem, students are given data[32] for the energy changes in the various steps of the ATP hydrolysis reaction catalyzed by myosin, which takes place in muscle cells to make muscle contraction possible. Students are asked to use what they know about chemical bonding from both the physics class and their prior chemistry experiences to explain the sign of the energy change at each step. Specifically, if the energy change is negative (corresponding to energy leaving the system), then bonds are being formed; if it is positive, bonds are being broken.

A second group problem-solving task, on temperature regulation, has the students reason about the signs of heat, work, and the change in internal energy of a system using the First Law of Thermodynamics, in a style similar to traditional physics problems.[33] However, the situations are biological, dealing with temperature regulation in mammals and other animals and leveraging students' knowledge of physiology (e.g. the difference between mammals, which maintain a constant internal temperature, and animals whose body temperatures depend on external conditions; the effects of metabolic reactions on thermal and chemical energy), and students are explicitly asked to separately analyze changes in thermal energy and chemical energy.

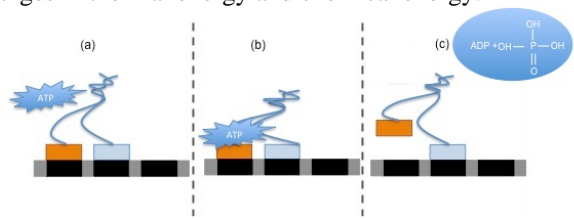

**FIG. 6.** The picture given to students in the kinesin task, along with a description of what is happening in each frame.

A third group problem-solving task deals with kinesin, a motor protein that "walks"[34] along microtubules to transport cargo within cells. This active transport is powered by the hydrolysis of ATP. Students are given a "frame-by-frame" description of the kinesin's motion (Figure 6), and in their groups produce energy bar charts[35] that account for the bonding between the kinesin and the microtubule, between the kinesin and the ATP, and the ATP hydrolysis reaction itself. This leads up to having the students discuss what it means to say that a cell "uses ATP to fuel molecular movement," producing more detailed explanations for phenomena they have encountered in biology on a more general level. The task is formulated in an open-ended way, and therefore there are many possible approaches the students can take in creating their energy bar charts (and we have in fact observed multiple approaches). They are explicitly asked to define their system, and are not told which objects to include as part of the system. They are also not told which energies to include in their bar charts, so student groups have taken different approaches about whether to use "chemical energy" or "potential energy," and whether to consider the chemical/potential energy "of" particular molecules, or of interactions among them. However, we would expect a correct solution to be internally consistent, with the total energy conserved in each frame, and the correct signs for the changes in energy associated with the formation and breaking of bonds. In many solution pathways, this means keeping track of energy conservation involving positive and negative energies.[36]

A culminating task for the chemical energy thread is an essay question (originally given on a midterm exam), shown in Fig. 7, that has students engage in interdisciplinary reconciliation around ATP hydrolysis.[37] As shown in the figure, the students are given two different representations: a potential energy diagram for a general chemical bond, and a chemical equation for this reaction showing the structure of each molecule. Students are asked to reconcile the idea (useful in biology) that the O-P bond in ATP is called a "high-energy bond"[38] because a large amount of energy is released when ATP is hydrolyzed, with the idea (based on modeling chemical bonds with potential energy) that an input of energy is required to break the bond. Successful reconciliation involves recognizing that both ideas are correct: the reaction includes both the breaking and formation of bonds, and the net effect is the release of energy.

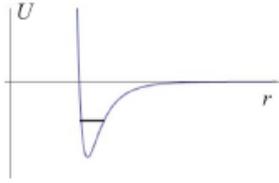

**FIG. 7.** The interdisciplinary reconciliation essay question on ATP hydrolysis.

## VI. EXAMPLES OF STUDENT OUTCOMES

Our evaluation of students developing ideas about chemical energy has been primarily qualitative and includes analyses of written student work, whole-class and small-group video data, and 48 semi-structured interviews with 23 students during the first two years of the NEXUS/Physics course. By focusing on qualitative descriptions of student thinking across the chemical energy thread we have begun to develop a picture of what an integrated understanding of chemical energy looks like. In this section we present examples of student data that illustrate the interdisciplinary reasoning about energy that is the intended outcome of the chemical energy thread. We then demonstrate how this descriptive data can be used to develop quantitative course-level assessments.

When the ATP essay question shown in Fig. 7 was given on an exam, students were asked to assess and reconcile the statements of "Justin," who says that energy is released when ATP is hydrolyzed, and "Kim," who claims based on a potential energy diagram that energy is required to break the phosphate bond in ATP. We present two exemplary student responses, from Jasper and Anya.

> *Jasper: Kim is right in her fundamental idea that it takes an input of energy to break bonds, even a weak one like the O-P bond. She inferred this from her PE graph based on the fact that if molecules are in the PE well, they are in a bound state. To escape the well, they must be "pushed out", which would require an input of energy. Justin is still right in the fact that hydrolyzing ATP releases energy, but this is because there are bonds being formed as well in the reaction, which acts to release energy. This is seen a bit easier in the molecule diagrams. What helps me think about PE problems is thinking of the gravitational analogy.*

*A ball at the edge of a table may have lots of PE, and if rolled off onto the ground, the PE converts to KE. The same is true for a bond. When a bond is formed, it is in a negative PE well, and KE must be released. To get bond out of the negative well back to 0, and positive input of KE is necessary to do so, hence why breaking bonds require and input of energy. The two's ideas can be reconciled, as they are both right.*

**Anya:** *Kim inferred this based on the fact that the bound state (the lowest point on the PE graph) has the point of lowest PE, and moving toward a non-bound state (aka, larger r/eventually breaking the O-P bond) corresponds to an increase in energy. This energy increase must come from somewhere according to the conservation of energy (can't just make it from nothing). In the end, both statements are correct – while it does require energy to sever the O-P bond, it is not much, and the ensuing energetic stability of the resulting ADP and $P_i$ molecules is much greater than when they were bound, resulting in a large energy release, much greater than the energy input required to break the bond.*

Both Jasper and Anya are able to leverage "physics" concepts about energy (e.g. conservation of energy, kinetic and potential energy) to explain this biochemical scenario. Jasper's response shows an ability to link bond energy and gravitational energy through an analogy as well as a coordinated use of representations (the PE graph and the molecule diagrams) to support his reasoning. Anya's response shows an ability to draw attention to the principle of conservation of energy in her discussion of bond breaking. While not all of the students made these connections (and while Anya's response is not complete since it is not clear from this response that new bonds are formed), these are the kinds of connections among energy concepts, representational forms, and epistemological frameworks that represent a desired outcome.

Analysis of interview data revealed students making connections beyond the specific prompts in the course. Betsy began an interview by spontaneously explaining an instance in which she saw the NEXUS/Physics course as helping her resolve an apparent contradiction between what she was learning in her chemistry and biology classes. In chemistry, she had learned that "it takes energy to break bonds, and when you form bonds you get energy back." Meanwhile, in her biology class, she had studied the difference between anabolic reactions, in which smaller molecules are built up into larger molecules, and catabolic reactions, in which larger molecules are broken down. Specifically, she had learned that catabolic reactions are needed in order to make anabolic reactions go, yet based on chemistry, she would have expected that anabolic reactions would release energy and catabolic reactions would require energy. Betsy began the process of reconciling these two principles with the specific case of ATP hydrolysis, which was supported by the chemical energy thread in the NEXUS/Physics course. As far as we can tell, Betsy made these connections on her own, since there was no explicit discussion of catabolic and anabolic reactions in the NEXUS/Physics course. While Betsy had not fully resolved this issue at the time of this interview, she demonstrated that she had identified a set of seemingly contradictory ideas and had begun to seek reconciliation. In addition to the specific content, Betsy experienced the physics class as creating opportunities to seek interdisciplinary coherence. She introduced the explanation of the chemical bond conundrum by reporting that "it feels like all of my classes are contradicting each other all the time, but the physics is kind of helping me pull it all together and understand that different things apply at different times." Betsy's ability to recognize variation in disciplinary frameworks and her desire to seek conceptual consistency across these frameworks illustrates the kind of outcome we hope this thread-based integrated curriculum can support.

We have drawn on this qualitative data to develop a strategy for evaluating students' evolving understanding of chemical energy at the class level. For a subset of tasks in the chemical energy thread we have developed formal rubrics.[6] This evaluation strategy gives us feedback about how students overall are understanding and linking the components of the thread. For example, the rubric we developed for analyzing the ATP essay question in Figure 7 assesses student responses along six dimensions: defining the reaction, energy in breaking/forming bonds, balance of energy, spontaneously generating connections between the potential energy curve and a physical picture, spontaneously generating connections to other concepts outside the problem, and coherence. (This goes beyond the standards by which students were graded on the actual exam; the "spontaneous" connections are those that were not explicitly required by the problem statement.) In the first year of the NEXUS/Physics course, we found that around half of the students (N=19) met or exceeded expectations on this question. While this result suggests that there is still work to be done, we cite this result to show that the examples from the interviews above are not outliers.

Our students have shown us that they are both interested and capable of coordinating ideas across their science courses, but it remains an ongoing challenge to design assessments that can both measure this development and productively inform new iterations of our curriculum. Our current approach, which is still in development, is to conduct a coordinated analysis of student progress across multiple rubrics and along multiple threads. This approach reflects our understanding of scientific expertise as involving integration and fluency of knowledge, not merely presence or absence of specific concepts.

## VII. CONCLUSION

A focus on chemical energy in the introductory physics course can help serve the needs of life sciences students by serving as a bridge between physics approaches to energy and the energy contexts most relevant to biology.[20] Unpacking chemical bond energy provides students with opportunities to reconcile seemingly contradictory ideas from the disciplines, and with a more coherent view of energy at different scales. Conceptualizing chemical energy as a thread means building up students' understanding of chemical energy by making explicit the links between different disciplinary ideas throughout the course.

The description of the chemical energy thread presented in this paper is a starting point, and will continue to be revised iteratively based on how students engage with it. Future directions include integrating chemical energy with our ongoing work on entropy and free energy[25], building conceptual links to coupled biochemical reactions (in which energy is not simply "released," but makes another reaction possible), and connecting chemical energy to optics through modeling photosynthesis. We invite the reader not to see our materials as a finished product that can be used anywhere, but to continue adapting them for new student populations and instructional settings. We also welcome the development of additional materials in this area and of other threads that support interdisciplinary connections.

## ACKNOWLEDGMENTS

The authors thank Chris Bauer, Melanie Cooper, Catherine Crouch, and Mike Klymkowsky for many valuable interdisciplinary conversations on chemical energy and thermodynamics that contributed significantly to this work. This work is supported by the NSF Graduate Research Fellowship (DGE 0750616), NSF-TUES DUE 11-22818, and the HHMI NEXUS grant.